\documentclass{article}
\usepackage{amsfonts}

\input{tcilatex}

\begin{document}

\title{Husimi operator and Husimi function for describing electron's
probability distribution in uniform magnetic field derived by virtue of the
entangled state representation \thanks{%
Work supported by the National Natural Science Foundation of China under
grant 10475056 } }
\author{$^{1,2,3}$Hong-yi Fan and $^{2,4}$Qin Guo \\
$^{1}$CCAST (World Laboratory), P O Box 8730, Beijing, 100080 China\\
$^{2}$Department of Physics, Shanghai Jiao Tong University. Shanghai 200030,
China\\
$^{3}$Department of Material Science and Engineering, University of Science\\
and Technology of China, Hefei, Anhui 230026, China\\
$^{4}$Department of Physics,Jiangxi Normal University, Nanchang 330022, China%
}
\maketitle

\begin{abstract}
For the first time we introduce the Husimi operator $\Delta _{h}\left(
\gamma ,\varepsilon ;\kappa \right) $ for studying Husimi distribution in
phase space $\left( \gamma ,\varepsilon \right) $ for electron's states in
uniform magnetic field, where $\kappa $ is the Gaussian spatial width
parameter. Using the Wigner operator in the entangled state $\left\langle
\lambda \right\vert $ representation [Hong-Yi Fan, Phys. Lett. A 301 (2002)
153; A 126 (1987) 145) we find that $\Delta _{h}\left( \gamma ,\varepsilon
;\kappa \right) $ is just a pure squeezed coherent state density operator $%
\left\vert \gamma ,\varepsilon \right\rangle _{\kappa \kappa }\left\langle
\gamma ,\varepsilon \right\vert ,$ which brings convenience for studying and
calculating the Husimi distribution. We in many ways demonstrate that the
Husimi distributions are Gaussian-broadened version of the Wigner
distributions. Throughout our calculation we have fully employed the
technique of integration within an ordered product of operators.

\end{abstract}

\section{Introduction}

Since the discovery of quantum Hall effect \cite{r1}-\cite{r4}, the motion
of an electron in the presence of magnetic field has brought an upsurge of
interest. The basic theory that underlies quantum Hall effect is the Landau
energy-level \cite{r5}-\cite{r6}. In Ref. \cite{r7} we have introduced an
entangled state representation $\left\vert \lambda \right\rangle $ to
describe this system which brings much convenience, for a review we refer to
Ref. \cite{r8}. This coincides with Dirac's guidance in Ref. \cite{r9}:"When
one has a particular problem to work out in quantum mechanics, one can
minimize the labor by using a representation in which the representatives of
the more important abstract quantities occurring in that problem are as
simple as possible". On the other hand, in quantum mechanics it is
impossible to specify simultaneously the position $Q\;$and the momentum $P$
of a particle due to Heisenberg uncertainty principle$.$ Thus Wigner's
quantum phase-space distribution theory \cite{r10}-\cite{r12} is of
increasing interest because it permits a direct comparison between classical
and quantum dynamics. Following the idea of gauge-invariant Wigner operator
proposed by Serimaa, Javanainen and Varro \cite{r13} we have constructed the
corresponding Wigner operator and Wigner function theory for electrons'
states in the $\left\vert \lambda \right\rangle $ representation in Ref.
\cite{r14}, as well as established the corresponding tomographic theory
which means the reconstruction of electron's Wigner distribution from the
tomographic data \cite{r15}. Let us briefly recall the original idea of
Wigner function. Feynman \cite{r16} summarized it as posing the following
question: If there is any density function $F_{w}\left( q,p\right) \;$in
quantum mechanics that satisfies \ \
\begin{equation}
\mathrm{P}(p)=\int_{-\infty }^{\infty }F_{w}\left( q,p\right) dq,\;\mathrm{P}%
\left( q\right) =\frac{1}{2\pi }\int_{-\infty }^{\infty }F_{w}\left(
q,p\right) dp,  \label{1}
\end{equation}%
where $P\left( q\right) \;\left[ P(p)\right] \;$is proportional to the
probability for finding the particle at $q$ [at $p$ in momentum space]. The
answer is
\begin{equation}
F_{w}\left( q,p\right) =Tr[\rho \triangle \left( q,p\right) ]=\frac{1}{2\pi }%
\int_{-\infty }^{\infty }\left\langle q+\frac{\upsilon }{2}\right\vert \rho
\left\vert q-\frac{\upsilon }{2}\right\rangle e^{-ip\upsilon }d\upsilon ,
\label{2}
\end{equation}%
where $\rho $ is a density operator, $\left\vert q\right\rangle $ is the
eigenvector of the coordinates operator, $Q\left\vert q\right\rangle
=q\left\vert q\right\rangle ,$ and $\triangle \left( q,p\right) $ is the
single-mode Wigner operator. In the coordinate representation $\triangle
\left( q,p\right) $ takes the form
\begin{eqnarray}
\triangle \left( q,p\right)  &=&\frac{1}{\left( 2\pi \right) ^{2}}\int
\int_{-\infty }^{\infty }dud\upsilon \exp \left[ iu\left( P-p\right)
+iv\left( Q-q\right) \right]   \nonumber \\
&=&\frac{1}{2\pi }\int_{-\infty }^{\infty }\left\vert q-\frac{\upsilon }{2}%
\right\rangle \left\langle q+\frac{\upsilon }{2}\right\vert e^{-ip\upsilon
}d\upsilon ,  \label{3}
\end{eqnarray}%
Eq. (\ref{1}) indicates that $\mathrm{P}\left( x\right) \;\left[ \mathrm{P}%
(p)\right] $ is the marginal distribution of $F_{w}(x,p).$ Using the
technique of integration within ordered product (IWOP) of operators \cite%
{r17}-\cite{r18}, we have performed the integral (\ref{3}) to obtain an
explicit operator \cite{r19}%
\begin{equation}
\triangle \left( q,p\right) =\frac{1}{\pi }:e^{-\left( q-Q\right)
^{2}-\left( p-P\right) ^{2}}:,  \label{4}
\end{equation}%
or%
\begin{equation}
\triangle \left( q,p\right) \rightarrow \triangle \left( \alpha ,\alpha
^{\ast }\right) =\frac{1}{\pi }:\exp \left[ -2\left( a^{\dagger }-\alpha
^{\ast }\right) \left( a-\alpha \right) \right] :,  \label{5}
\end{equation}%
where $\alpha =\left( q+ip\right) /\sqrt{2}$, $:$ $:$ means normal ordering
symbol, $Q=\left( a+a^{\dagger }\right) /\sqrt{2},$ $P=\left( a-a^{\dagger
}\right) /\left( i\sqrt{2}\right) $\ is the momentum operator whose
eigenvector is $\left\vert p\right\rangle $. It then follows from (\ref{4})
that one-sided integral over the Wigner operator yields the pure position
state density operator
\begin{equation}
\int_{-\infty }^{\infty }dp\triangle \left( q,p\right) =\frac{1}{\sqrt{\pi }}%
:e^{-\left( q-Q\right) ^{2}}:=\left\vert q\right\rangle \left\langle
q\right\vert ,  \label{6}
\end{equation}%
and pure momentum state density operator
\begin{equation}
\int_{-\infty }^{\infty }dq\triangle \left( q,p\right) =\frac{1}{\sqrt{\pi }}%
:e^{-\left( p-P\right) ^{2}}:=\left\vert p\right\rangle \left\langle
p\right\vert ,  \label{7}
\end{equation}%
respectively, so the marginal distribution of the Wigner function is $%
\int_{-\infty }^{\infty }dp\left\langle \psi \right\vert \triangle \left(
q,p\right) \left\vert \psi \right\rangle =|\psi \left( q\right) |^{2}$ or $%
\int_{-\infty }^{\infty }dq\left\langle \psi \right\vert \triangle \left(
q,p\right) \left\vert \psi \right\rangle =|\psi \left( p\right) |^{2},$
respectively. However, as many authors have pointed out that the Wigner
function $F_{w}\left( q,p\right) $ is not a probability distribution since
it may takes on both positive and negative values. To quickly see this we
can use $D\left( \alpha \right) =\exp \left( \alpha a^{\dagger }-\alpha
^{\ast }a\right) ,$ $N=a^{\dagger }a,$ to express (\ref{5}) as \ $\triangle
\left( \alpha ,\alpha ^{\ast }\right) =\frac{1}{\pi }D\left( \alpha \right)
\left( -1\right) ^{N}D^{\dagger }\left( \alpha \right) .$ Let $D^{\dagger
}\left( \alpha \right) \left\vert \psi \right\rangle =\left\vert \phi
\right\rangle ,$ then from

\begin{eqnarray}
\left\langle \psi \right\vert \triangle \left( \alpha ,\alpha ^{\ast
}\right) \left\vert \psi \right\rangle &=&\left\langle \phi \right\vert
\left( -1\right) ^{N}\left\vert \phi \right\rangle  \nonumber \\
&=&\left\langle \phi \right\vert \sum\limits_{n^{\prime }=0}^{\infty
}\left\vert n^{\prime }\right\rangle \left\langle n^{\prime }\right\vert
\left( -1\right) ^{N}\sum\limits_{n=0}^{\infty }\left\vert n\right\rangle
\left\langle n\right\vert \left. \phi \right\rangle
=\sum\limits_{n=0}^{\infty }\left( -1\right) ^{n}\left\vert \left\langle
n\right\vert \left. \phi \right\rangle \right\vert ^{2},  \label{8}
\end{eqnarray}%
where the existence of $\left( -1\right) ^{n}$ implies that the Wigner
distribution function itself is not a probability distribution due to $%
\left( -1\right) ^{n}$ being both positive and negative. To overcome this
shortcomings, the so-called Husimi distribution function $F_{h}\left(
q,p,\kappa \right) $ is introduced \cite{r20}, which is defined in a manner
that guarantees it to be non-negative and gives it a probability
interpretation. Its definition is smoothing out the Wigner function by
averaging over a "coarse graining" function,
\begin{equation}
F_{h}\left( q,p,s\right) =\int \int_{-\infty }^{\infty }dq^{\prime
}dp^{\prime }F_{w}\left( q^{\prime },p^{\prime }\right) \exp \left[ -s\left(
q^{\prime }-q\right) ^{2}-\frac{\left( p^{\prime }-p\right) ^{2}}{s}\right] ,
\label{9}
\end{equation}%
where $s$ is the Gaussian spatial width parameter, which determines the
relative resolution in $p$-space versus $q$-space but is free to be chosen.
It is understood that the Husimi density is given by the projection of the
wave function $\psi $ onto coherent states localized in phase space $\left(
p,q\right) $ with a minimum product of the uncertainties $\Delta P=\sqrt{%
\frac{s\hbar }{2}},$ $\Delta Q=\sqrt{\frac{\hbar }{2s}}$. In this sense $s$
plays the role of squeezing-parameter. In Refs. \cite{r21}-\cite{r22} the
Husimi operator which corresponds to Husimi function is introduced, which
turns out to be a pure squeezed coherent state projector. An interesting
question thus naturally arises: how to introduce Husimi functions of phase
space for describing probability distribution\ of electron states in uniform
magnetic field (UMF)? To our knowledge, such a question has not been posed
in the literature before. As emphasized by Serimaa, Javanainen and Varro
\cite{r13} that when one wants to establish phase space distribution theory
for electron moving in UMF with the gauge potential $\mathbf{\vec{A}}=\left(
-\frac{1}{2}By,\frac{1}{2}Bx,0\right) ,$ electron's canonical momentum
operators $\left( p_{x},p_{y}\right) $ (conjugate to electron's coordinate
operator $x,y$) should be replaced by its gauge-invariant kinetic momentum
(in the units of $\hbar =c=1,$ $c$ denotes the speed of light), $\Pi
_{x}=p_{x}+eA_{x},\;\Pi _{y}=p_{y}+eA_{y}.$ Correspondingly, the Wigner
operator for describing electrons' motion in UMF should involve $\Pi _{x}\ $%
and $\Pi _{y}$ as ingredient operators and therefore is gauge invariant. In
Ref. \cite{r14} we have proposed Wigner operator in the entangled state
representation (i.e. electron's position representation, denoted by $%
\left\vert \lambda \right\rangle $). In this work we shall first introduce
the Husimi operator $\Delta _{h}\left( \varepsilon ,\gamma ;\kappa \right) $
by using this Wigner operator.\ Remarkably, as one can see shortly later,
that the Husimi operator $\Delta _{h}\left( \varepsilon ,\gamma ;\kappa
\right) $ is just a pure squeezed coherent state density operator $%
\left\vert \varepsilon ,\gamma \right\rangle _{\kappa \kappa }\left\langle
\varepsilon ,\gamma \right\vert ,$ (the explicit form of $\left\vert
\varepsilon ,\gamma \right\rangle _{\kappa }$ in Fock space can also be
deduced, see Eq. (\ref{41}) below), which brings much convenience to
studying Husimi functions for various electron's states. Thus a phase space
Husimi distribution theory for electron moving in uniform magnetic field
(UMF) can be successfully established. The work is arranged as follows: In
Sec. 2 we briefly review the concise features of the normally ordered form
of gauge invariant Wigner operator $\Delta _{B}\left( \gamma ,\varepsilon
\right) $ in expressing the marginal distribution probability in the $%
\left\vert \lambda \right\rangle $ representation and its conjugate
representation $\left\vert \zeta \right\rangle $ (electron's canonical
momentum representation). In Sec. 3 we first introduce the Husimi operator $%
\Delta _{h}\left( \varepsilon ,\gamma ;\kappa \right) $ and then derive its
normally ordered form, correspondingly, we introduce Husimi function for
describing electron's probability distribution. The marginal distributions
of Husimi function turns out to be Gaussian-broadened version of the Wigner
marginal distributions. We also notice that the Gaussian spatial width
parameter can be related to the intensity of magnetic field. In Sec. 4 we
introduce the two-mode squeezed coherent state $\left\vert \gamma
,\varepsilon \right\rangle _{\kappa }$ and show its capability of
constituting a quantum mechanical representation, we then find that the pure
state $\left\vert \gamma ,\varepsilon \right\rangle _{\kappa \kappa
}\left\langle \gamma ,\varepsilon \right\vert $ is just the Husimi operator$%
, $ so $\left\vert \gamma ,\varepsilon \right\rangle _{\kappa }$ is a good
representation for illustrating the Husimi function. In Sec. 5 we further
analyze physical explanation of Husimi function of electron's states by
calculating the uncertainty relation of electron's position and momentum. In
Sec. 6 we calculate the Husimi function of various electron's states in a
concise and neat way. In Sec. 7 we discuss squeezing of Husimi function by
variation of magnetic field. In so doing, the Husimi function theory for
describing distribution\ of electron states in uniform magnetic field is
established and the relationship between Husimi function and Wigner function
is clearly illuminated.

\section{Wigner operator in entangled state representation and its marginal
distributions}

\bigskip The Hamiltonian for electron in UMF is $H=\left( \Pi _{+}\Pi _{-}+%
\frac{1}{2}\right) \Omega ,$ the ladder operators are related to electron's
kinetic momenta $\left( \Pi _{x},\Pi _{y}\right) $,$\;\Pi _{\pm }=\frac{\Pi
_{x}\pm i\Pi _{y}}{\sqrt{2M\Omega }},$ $\Omega =\frac{eB}{M}$\ is the
cyclotron frequency, $M\;$is the mass of electron. For the appropriate
gauge-invariant Wigner operator \cite{r13}%
\begin{eqnarray}
\Delta _{B}\left( \vec{k},\vec{q}\right) &=&\frac{1}{\left( 2\pi \right) ^{4}%
}\int \int_{-\infty }^{\infty }d^{2}ud^{2}\upsilon \exp \left[ iu\left( \vec{%
\Pi}-\vec{k}\right) +iv\left( \vec{Q}-\vec{q}\right) \right] ,  \label{10} \\
\text{where \ }\vec{k} &=&\left( k_{1},k_{2}\right) ,\text{ }\vec{q}=\left(
q_{1},q_{2}\right) ,\text{ }\vec{\Pi}=\left( \Pi _{x},\Pi _{y}\right) ,\text{
}\vec{Q}=\left( x,y\right) ,  \nonumber
\end{eqnarray}%
we have proved in Ref. \cite{r14} that $\Delta _{B}\left( \vec{k},\vec{q}%
\right) $ in the entangled state representation $\left\vert \lambda
\right\rangle $ \cite{r7}-\cite{r8} is expressed as (somehow similar in form
to (\ref{3}))%
\begin{eqnarray}
\Delta _{B}\left( \gamma ,\varepsilon \right) &=&\int \frac{d^{2}\lambda }{%
\pi ^{3}}\left\vert \varepsilon ^{\ast }-\lambda \right\rangle \left\langle
\varepsilon ^{\ast }+\lambda \right\vert e^{\gamma ^{\ast }\lambda ^{\ast
}-\gamma \lambda },  \label{11} \\
\text{where \ }\gamma &=&\chi +i\sigma ^{\ast },\text{ }\varepsilon =\chi
-i\sigma ^{\ast },  \nonumber \\
\chi &=&\sqrt{\frac{M\Omega }{2}}\left( q_{1}+iq_{2}\right) +i\sqrt{\frac{1}{%
2M\Omega }}\left( k_{1}+ik_{2}\right) ,\text{ }\sigma =\sqrt{\frac{1}{%
2M\Omega }}\left( k_{1}-ik_{2}\right) ,  \nonumber
\end{eqnarray}%
the state $\left\vert \lambda \right\rangle $ is
\begin{equation}
\left\vert \lambda \right\rangle =\exp \left[ -\frac{1}{2}\left\vert \lambda
\right\vert ^{2}-i\lambda \Pi _{+}+\lambda ^{\ast }K_{+}+i\Pi _{+}K_{+}%
\right] \left\vert 00\right\rangle ,\;\lambda =\lambda _{1}+i\lambda _{2},
\label{12}
\end{equation}%
here the vacuum state is annihilated by $\Pi _{-}\left\vert 00\right\rangle
=0,\;K_{-}\left\vert 00\right\rangle =0,$ $K_{\pm }$ are linear combination
of guiding centers $x_{0}$ and $y_{0\;}$\cite{r6}\cite{r25},
\begin{equation}
K_{\pm }=\sqrt{\frac{M\Omega }{2}}\left( x_{0}\mp iy_{0}\right) ,\;\;
\label{13}
\end{equation}%
\begin{equation}
x_{0}=x-\frac{\Pi _{y}}{M\Omega },\;\;y_{0}=y+\frac{\Pi _{x}}{M\Omega }%
.\;\;\;  \label{14}
\end{equation}%
Note that the above operators obey commutative relations,
\begin{eqnarray}
\left[ \Pi _{-},\Pi _{+}\right] &=&1,\;\left[ K_{-},K_{+}\right] =1,\;
\label{15} \\
\left[ K_{\pm },\Pi _{\pm }\right] &=&0,\;\left[ x_{0},\Pi _{\pm }\right]
=0,\;\left[ y_{0},\Pi _{\pm }\right] =0,\;[x,y]=0,  \nonumber \\
\left[ x_{0},y_{0}\right] &=&\frac{i}{M\Omega },\;\left[ \Pi _{x},\Pi _{y}%
\right] =-iM\Omega ,  \nonumber
\end{eqnarray}%
$\left\vert \lambda \right\rangle $ is named entangled state \cite{r15}. The
motivation of introducing $\left\vert \lambda \right\rangle $ lies in two
aspects: Firstly, when magnetic field $\mathbf{\vec{B}}$ applies what we
have operators physically describing the system at hand are the guiding
centers and kinetic momenta. In other words, the dynamic variables in the
Hamiltonian are $\Pi _{_{\pm }}$, so the corresponding position eigenvector
should be expressed by $\Pi _{\pm }$\ as well as $K_{\pm }$. Secondly, $%
\left\vert \lambda \right\rangle $ can conveniently describe the position of
an electron in a uniform magnetic field, i.e. $\left\vert \lambda
\right\rangle $ satisfies the coordinate eigenvector equation
\begin{equation}
\left( K_{+}+i\Pi _{-}\right) \left\vert \lambda \right\rangle =\lambda
\left\vert \lambda \right\rangle ,\;\;\left( K_{-}-i\Pi _{+}\right)
\left\vert \lambda \right\rangle =\lambda ^{\ast }\left\vert \lambda
\right\rangle .  \label{16}
\end{equation}%
Combining (\ref{12})-(\ref{16}) yields
\begin{equation}
x=\sqrt{\frac{1}{2M\Omega }}\left( K_{+}+K_{-}-i\Pi _{+}+i\Pi _{-}\right) ,\
y=\frac{i}{\sqrt{2M\Omega }}\left( K_{+}-K_{-}+i\Pi _{+}+i\Pi _{-}\right) ,
\label{17}
\end{equation}%
\begin{equation}
x\left\vert \lambda \right\rangle =\sqrt{\frac{2}{M\Omega }}\lambda
_{1}\left\vert \lambda \right\rangle ,\;\;y\left\vert \lambda \right\rangle
=-\sqrt{\frac{2}{M\Omega }}\lambda _{2}\left\vert \lambda \right\rangle .
\label{18}
\end{equation}%
Moreover, the Wigner operator expressed by (\ref{11}) in $\left\vert \lambda
\right\rangle $ representation automatically includes the contribution form
the magnetic field, this is another merit of introducing $\left\vert \lambda
\right\rangle $. The advantage of $\Delta _{B}\left( \gamma ,\varepsilon
\right) $ also lies in that from (\ref{11}) we can easily derive its
marginal distributions. In fact, using the normally ordered form of $%
\left\vert 00\right\rangle \left\langle 00\right\vert =:\exp \left[ -\Pi
_{+}\Pi _{-}-K_{+}K_{-}\right] :$ and the IWOP technique \cite{r17}-\cite%
{r18} we can perform the integration in (\ref{11}) to derive the normally
ordered form of the Wigner operator $\Delta _{B}\left( \gamma ,\varepsilon
\right) $

\begin{eqnarray}
&&\Delta _{B}\left( \gamma ,\varepsilon \right)  \nonumber \\
&=&\int \frac{d^{2}\lambda }{\pi ^{3}}\colon \exp \{-|\varepsilon ^{\ast
}|^{2}-|\lambda |^{2}-i\left( \varepsilon ^{\ast }-\lambda \right) \Pi
_{+}+\left( \varepsilon -\lambda ^{\ast }\right) K_{+}+i\left( \varepsilon
+\lambda ^{\ast }\right) \Pi _{-}  \nonumber \\
&&+\left( \varepsilon ^{\ast }+\lambda \right) K_{-}+i\Pi _{+}K_{+}-i\Pi
_{-}K_{-}-\Pi _{+}\Pi _{-}-K_{+}K_{-}+\gamma ^{\ast }\lambda ^{\ast }-\gamma
\lambda \}\colon  \nonumber \\
&=&\frac{1}{\pi ^{2}}\colon \exp \{-\left[ \varepsilon ^{\ast }-\left(
K_{+}+i\Pi _{-}\right) \right] \left[ \varepsilon -\left( K_{-}-i\Pi
_{+}\right) \right]  \nonumber \\
&&-\left[ \gamma ^{\ast }-\left( K_{+}-i\Pi _{-}\right) \right] \left[
\gamma -\left( K_{-}+i\Pi _{+}\right) \right] \}\colon .  \label{19}
\end{eqnarray}%
As (\ref{11}) indicates, $\chi =\frac{1}{2}\left( \gamma +\varepsilon
\right) ,$ $\sigma ^{\ast }=\frac{1}{2i}\left( \gamma -\varepsilon \right) ,$
then (\ref{19}) becomes%
\begin{equation}
\Delta _{B}\left( \gamma ,\varepsilon \right) =\frac{1}{\pi ^{2}}\colon \exp
\{-2\left( K_{+}-\chi ^{\ast }\right) \left( K_{-}-\chi \right) -2\left( \Pi
_{+}-\sigma ^{\ast }\right) \left( \Pi _{-}-\sigma \right) \}\colon ,
\label{20}
\end{equation}%
which is a 2-dimensional generalization of Eq. (\ref{5}), so (\ref{11}) is a
correct choice. Note\ that the normally ordered form of the projector $%
\left\vert \lambda \right\rangle \left\langle \lambda \right\vert $ is
\begin{equation}
\left\vert \lambda \right\rangle \left\langle \lambda \right\vert =\colon
\exp \{-\left[ \lambda ^{\ast }-\left( K_{-}-i\Pi _{+}\right) \right] \left[
\lambda -\left( K_{+}+i\Pi _{-}\right) \right] \}\colon ,  \label{21}
\end{equation}%
with the completeness $\int \frac{d^{2}\lambda }{\pi }\left\vert \lambda
\right\rangle \left\langle \lambda \right\vert =1,$ so integrating (\ref{19}%
) over $d^{2}\gamma $ and using (\ref{21}) we see

\begin{eqnarray}
\pi \left\langle \psi \right\vert \int d^{2}\gamma \Delta _{B}\left( \gamma
,\varepsilon \right) \left\vert \psi \right\rangle &=&\colon \exp \{-\left[
\varepsilon ^{\ast }-\left( K_{+}+i\Pi _{-}\right) \right] \left[
\varepsilon -\left( K_{-}-i\Pi _{+}\right) \right] \}\colon  \nonumber \\
\text{\ } &=&\left\langle \psi \right. \left\vert \lambda \right\rangle
\left\langle \lambda \right\vert |_{\lambda =\varepsilon ^{\ast }}\left\vert
\psi \right\rangle =\left\vert \left\langle \psi \right. \left\vert \lambda
\right\rangle \right\vert ^{2}|_{\lambda =\varepsilon ^{\ast }}.  \label{22}
\end{eqnarray}

$\left\vert \left\langle \lambda \right\vert \left. \psi \right\rangle
\right\vert ^{2}$ is proportional to the probability for finding the
electron with position value $\left[ \text{ }\sqrt{\frac{2}{M\Omega }}%
\lambda _{1},-\sqrt{\frac{2}{M\Omega }}\lambda _{2}\right] $. Note $%
\left\langle \lambda \right\vert \left. \lambda ^{\prime }\right\rangle =\pi
\delta \left( \lambda -\lambda ^{\prime }\right) \delta \left( \lambda
^{\ast }-\lambda ^{\prime \ast }\right) \equiv \pi \delta ^{\left( 2\right)
}\left( \lambda -\lambda ^{\prime }\right) .$ On the other hand, integrating
(\ref{19}) over $d^{2}\varepsilon $ leads to%
\begin{eqnarray}
\pi \left\langle \psi \right\vert \int d^{2}\varepsilon \Delta _{B}\left(
\gamma ,\varepsilon \right) \left\vert \psi \right\rangle &=&\colon \exp \{-
\left[ \gamma ^{\ast }-\left( K_{+}-i\Pi _{-}\right) \right] \left[ \gamma
-\left( i\Pi _{+}+K_{-}\right) \right] \}\colon  \label{23} \\
&=&\left\langle \psi \right. \left\vert \zeta \right\rangle \left\langle
\zeta \right\vert |_{\zeta =-\gamma ^{\ast }}\left\vert \psi \right\rangle
=\left\vert \left\langle \psi \right. \left\vert \zeta \right\rangle
\right\vert ^{2}|_{\zeta =-\gamma ^{\ast }},  \nonumber
\end{eqnarray}%
where we have defined the state vector $\left\vert \zeta \right\rangle $ as
\begin{equation}
\left\vert \zeta \right\rangle =\exp \left[ -\frac{1}{2}\left\vert \zeta
\right\vert ^{2}-i\zeta \Pi _{+}-\zeta ^{\ast }K_{+}-i\Pi _{+}K_{+}\right]
\left\vert 00\right\rangle ,\text{ }\zeta =\zeta _{1}+i\zeta _{2},
\label{24}
\end{equation}%
and

\begin{eqnarray}
\left\vert \zeta \right\rangle \left\langle \zeta \right\vert &=&\colon \exp
\{-\left\vert \zeta \right\vert ^{2}-i\zeta \Pi _{+}-\zeta ^{\ast
}K_{+}-i\Pi _{+}K_{+}  \nonumber \\
&&+i\zeta ^{\ast }\Pi _{-}-\zeta K_{-}+i\Pi _{-}K_{-}-\Pi _{+}\Pi
_{-}-K_{+}K_{-}\}\colon  \nonumber \\
&=&\colon \exp \{-\left[ \zeta -\left( i\Pi _{-}-K_{+}\right) \right] \left[
\zeta ^{\ast }-\left( -i\Pi _{+}-K_{-}\right) \right] \}\colon ,  \label{25}
\end{eqnarray}

with the completeness $\int \frac{d^{2}\zeta }{\pi }\left\vert \zeta
\right\rangle \left\langle \zeta \right\vert =1.$ $\left\vert \zeta
\right\rangle $ is the common eigenvector of the canonical momenta ($%
P_{x},P_{y}),$ which can be shown as the following. In fact, due to%
\begin{equation}
\left( i\Pi _{-}-K_{+}\right) \left\vert \zeta \right\rangle =\zeta
\left\vert \zeta \right\rangle ,\text{ \ }\left( K_{-}+i\Pi _{+}\right)
\left\vert \zeta \right\rangle =-\zeta ^{\ast }\left\vert \zeta \right\rangle
\label{26}
\end{equation}%
and using$\;\;\;$%
\begin{eqnarray}
p_{x} &=&\sqrt{\frac{M\Omega }{8}}\left[ \Pi _{+}+\Pi _{-}+iK_{+}-iK_{-}%
\right] =\frac{\Pi _{x}}{2}+\frac{M\Omega }{2}y_{0},  \label{27} \\
p_{y} &=&\sqrt{\frac{M\Omega }{8}}\left[ i\Pi _{-}-i\Pi _{+}-K_{+}-K_{-}%
\right] =\frac{\Pi _{y}}{2}-\frac{M\Omega }{2}x_{0},  \nonumber
\end{eqnarray}%
we see%
\begin{equation}
p_{x}\left\vert \zeta \right\rangle =\sqrt{\frac{M\Omega }{2}}\zeta
_{2}\left\vert \zeta \right\rangle ,\;\text{\ }p_{y}\left\vert \zeta
\right\rangle =\sqrt{\frac{M\Omega }{2}}\zeta _{1}\left\vert \zeta
\right\rangle .  \label{28}
\end{equation}%
Thus $\left\vert \left\langle \psi \right\vert \left. \zeta \right\rangle
\right\vert ^{2}$ in (\ref{23}) is proportional to the probability for
finding the electron with momentum value ($\sqrt{\frac{M\Omega }{2}}\zeta
_{2},\sqrt{\frac{M\Omega }{2}}\zeta _{1})$. Combine (\ref{22}) and (\ref{23}%
) we see that the marginal distributions of the Wigner function for electron
states are physical meaningful in the entangled state representation $%
\left\vert \lambda \right\rangle $ (or $\left\vert \zeta \right\rangle $).
This in turn explains that the Wigner operator $\Delta _{B}\left( \gamma
,\varepsilon \right) $ expressed in $\left\langle \lambda \right\vert $
representation is a convenient choice which possesses the correct
statistical meaning. Note%
\begin{equation}
\int d^{2}\varepsilon \int d^{2}\gamma \Delta _{B}\left( \gamma ,\varepsilon
\right) =1.  \label{29}
\end{equation}%
For a general theory of entangled Wigner function we refer to \cite{r23}.

\section{Husimi operator: normally ordered form; the marginal distributions
of Husimi distribution function}

In this section we want to introduce the Husimi function $W_{h}\left( \gamma
,\varepsilon ;k\right) $ for describing electron's probability distribution,
the corresponding Husimi operator $\Delta _{h}\left( \gamma ,\varepsilon
;k\right) $, in reference to Eq. (\ref{9}), is defined as smoothing out $%
\Delta _{B}\left( \gamma ^{\prime },\varepsilon ^{\prime }\right) $ by
averaging over a "coarse graining" function,

\begin{equation}
\Delta _{h}\left( \gamma ,\varepsilon ;k\right) =4\int d^{2}\gamma ^{\prime
}d^{2}\varepsilon ^{\prime }\Delta _{B}\left( \gamma ^{\prime },\varepsilon
^{\prime }\right) \exp \left[ -\kappa \left\vert \varepsilon -\varepsilon
^{\prime }\right\vert ^{2}-\frac{\left\vert \gamma -\gamma ^{\prime
}\right\vert ^{2}}{\kappa }\right] ,  \label{30}
\end{equation}%
where $\kappa $ is the Gaussian spatial width parameter, which is free to be
chosen, and $W_{h}\left( \gamma ,\varepsilon ;k\right) =\left\langle \psi
\right\vert \Delta _{h}\left( \gamma ,\varepsilon ,\kappa \right) \left\vert
\psi \right\rangle $. Using (\ref{19}) and the IWOP technique we perform the
integration in (\ref{30}),

\begin{eqnarray}
\Delta _{h}\left( \gamma ,\varepsilon ;k\right) &=&\frac{4}{\pi ^{2}}\int
d^{2}\gamma ^{\prime }d^{2}\varepsilon ^{\prime }\colon \exp \{-\left[
\varepsilon ^{\prime \ast }-\left( K_{+}+i\Pi _{-}\right) \right] \left[
\varepsilon ^{\prime }-\left( K_{-}-i\Pi _{+}\right) \right]  \nonumber \\
&&-\left[ \gamma ^{\prime \ast }-\left( K_{+}-i\Pi _{-}\right) \right] \left[
\gamma ^{\prime }-\left( K_{-}+i\Pi _{+}\right) \right] \}\colon \exp
[-\kappa \left\vert \varepsilon -\varepsilon ^{\prime }\right\vert ^{2}-%
\frac{\left\vert \gamma -\gamma ^{\prime }\right\vert ^{2}}{\kappa }]
\nonumber \\
&=&\frac{4\kappa }{\left( 1+\kappa \right) ^{2}}\colon \exp \{-\frac{\kappa
}{1+\kappa }\left[ \varepsilon ^{\ast }-\left( K_{+}+i\Pi _{-}\right) \right]
\left[ \varepsilon -\left( K_{-}-i\Pi _{+}\right) \right]  \nonumber \\
&&-\frac{1}{1+\kappa }\left[ \gamma ^{\ast }-\left( K_{+}-i\Pi _{-}\right) %
\right] \left[ \gamma -\left( K_{-}+i\Pi _{+}\right) \right] \ \}\colon ,
\label{31}
\end{eqnarray}%
which is the explicit normally ordered form of the Husimi operator. Using $%
\gamma =\gamma _{1}+i\gamma _{2},$\ $\varepsilon =\varepsilon
_{1}+i\varepsilon _{2},$ (\ref{17}) and (\ref{27}) we can further change (%
\ref{31}) into the form

\begin{eqnarray}
\Delta _{h}\left( \gamma ,\varepsilon ;k\right) &=&\frac{4\kappa }{\left(
1+\kappa \right) ^{2}}\colon \exp \{-\frac{\kappa }{1+\kappa }\left[ \left(
\varepsilon _{1}-\sqrt{\frac{M\Omega }{2}}x\right) ^{2}+\left( \varepsilon
_{2}-\sqrt{\frac{M\Omega }{2}}y\right) ^{2}\right]  \nonumber \\
&&-\frac{1}{1+\kappa }\left[ \left( \gamma _{1}+\sqrt{\frac{2}{M\Omega }}%
p_{y}\right) ^{2}+\left( \gamma _{2}-\sqrt{\frac{2}{M\Omega }}p_{x}\right)
^{2}\right] \}\colon ,  \label{32}
\end{eqnarray}%
Using (\ref{31}) we perform the one-sided integration $d^{2}\gamma $ over $%
\Delta _{h},$

\begin{eqnarray}
&&\int \frac{d^{2}\gamma }{\pi }\Delta _{h}\left( \gamma ,\varepsilon
;k\right)  \label{33} \\
&=&\frac{4\kappa }{1+\kappa }\colon \exp \{\frac{-\kappa }{1+\kappa }\left(
\varepsilon ^{\ast }-K_{+}-i\Pi _{-}\right) \left( \varepsilon -K_{-}+i\Pi
_{+}\right) \}\colon .  \nonumber
\end{eqnarray}%
On the other hand, using the $\left\vert \lambda \right\rangle $
representation in (\ref{21}) and $x\left\vert \lambda \right\rangle =\sqrt{%
\frac{2}{M\Omega }}\lambda _{1}\left\vert \lambda \right\rangle
,\;y\left\vert \lambda \right\rangle =-\sqrt{\frac{2}{M\Omega }}\lambda
_{2}\left\vert \lambda \right\rangle $ in (\ref{18}) as well as the IWOP
technique we can derive the operator identity

\begin{eqnarray}
&&\exp \{g\left[ \left( s_{1}-\sqrt{\frac{M\Omega }{2}}x\right) ^{2}+\left(
s_{2}-\sqrt{\frac{M\Omega }{2}}y\right) ^{2}\right] \}  \nonumber \\
&=&\int \frac{d^{2}\lambda }{\pi }\exp \{g\left[ \left( s_{1}-\lambda
_{1}\right) ^{2}+\left( s_{2}-\lambda _{2}\right) ^{2}\right] \}\left\vert
\lambda \right\rangle \left\langle \lambda \right\vert  \nonumber \\
&=&\int \frac{d^{2}\lambda }{\pi }\colon \exp \{-\left( 1-g\right)
\left\vert \lambda \right\vert ^{2}+\lambda \left( K_{-}-i\Pi _{+}-gs\right)
+\lambda ^{\ast }\left( K_{+}+i\Pi _{-}-gs^{\ast }\right)  \nonumber \\
&&+g|s|^{2}-\left( K_{-}-i\Pi _{+}\right) \left( K_{+}+i\Pi _{-}\right)
\}\colon  \nonumber \\
&=&\frac{1}{1-g}\colon \exp \{\frac{g}{1-g}\left( s^{\ast }-K_{+}-i\Pi
_{-}\right) \left( s-K_{-}+i\Pi _{+}\right) \}\colon ,  \label{34}
\end{eqnarray}%
\ where $s=s_{1}+is_{2}.$ So (\ref{33}) can be simplified as (identifying $%
-\kappa $ in (\ref{33}) as $g$ in (\ref{34})$)$

\begin{equation}
\int \frac{d^{2}\gamma }{\pi }\Delta _{h}\left( \gamma ,\varepsilon ;\kappa
\right) =4\kappa e^{-\kappa \left[ \left( \varepsilon _{1}-\sqrt{\frac{%
M\Omega }{2}}x\right) ^{2}+\left( \varepsilon _{2}-\sqrt{\frac{M\Omega }{2}}%
y\right) ^{2}\right] },  \label{35}
\end{equation}%
thus the marginal distribution of Husimi operator is a Gaussian operator
with the factor $\kappa $. It then follows from (\ref{35}), (\ref{22}) and (%
\ref{18}) the marginal distribution of Husimi function in "$\lambda -$%
direction",%
\begin{eqnarray}
\int \frac{d^{2}\gamma }{\pi }W_{h}\left( \gamma ,\varepsilon ;k\right)
&=&\left\langle \psi \right\vert \int \frac{d^{2}\gamma }{\pi }\Delta
_{h}\left( \gamma ,\varepsilon ;\kappa \right) \left\vert \psi \right\rangle
\label{36} \\
&=&4\kappa \left\langle \psi \right\vert \int \frac{d^{2}\lambda }{\pi }%
e^{-\kappa \left[ \left( \varepsilon _{1}-\sqrt{\frac{M\Omega }{2}}x\right)
^{2}+\left( \varepsilon _{2}-\sqrt{\frac{M\Omega }{2}}y\right) ^{2}\right]
}\left\vert \lambda \right\rangle \left\langle \lambda \right\vert \left.
\psi \right\rangle  \nonumber \\
&=&4\kappa \left\langle \psi \right\vert \int \frac{d^{2}\lambda }{\pi }%
e^{^{^{-\kappa \left[ \left( \varepsilon _{1}-\lambda _{1}\right)
^{2}+\left( \varepsilon _{2}+\lambda _{2}\right) ^{2}\right] }}}\left\vert
\lambda \right\rangle \left\langle \lambda \right\vert \left. \psi
\right\rangle  \nonumber \\
&=&4\kappa \int \frac{d^{2}\lambda }{\pi }e^{^{-\kappa \left\vert
\varepsilon -\lambda ^{\ast }\right\vert ^{2}}}|\psi \left( \lambda \right)
|^{2}.  \nonumber
\end{eqnarray}%
Comparing (\ref{36}) with (\ref{22}) we see that (\ref{36}) is a
Gaussian-broadened version of the quantal position probability distribution $%
|\psi \left( \lambda \right) |^{2}$ (one marginal distribution of the Wigner
function)$.$ Similarly, performing the one-sided integration $%
d^{2}\varepsilon $ over $\Delta _{h}$ in (\ref{32}) leads to%
\begin{eqnarray}
&&\int \frac{d^{2}\varepsilon }{\pi }\Delta _{h}\left( \gamma ,\varepsilon
;\kappa \right)  \label{37} \\
&=&\frac{4}{1+\kappa }:\exp \{-\frac{1}{1+\kappa }\left[ \gamma ^{\ast
}-\left( K_{+}-i\Pi _{-}\right) \right] \left[ \gamma -\left( i\Pi
_{+}+K_{-}\right) \right] \ \}:.  \nonumber
\end{eqnarray}%
From (\ref{25}) and (\ref{28}) as well as the IWOP technique we can prove
another operator identity

\begin{eqnarray}
&&\exp \{g\left[ \left( v_{1}+\sqrt{\frac{2}{M\Omega }}p_{y}\right)
^{2}+\left( v_{2}-\sqrt{\frac{2}{M\Omega }}p_{x}\right) ^{2}\right] \}
\nonumber \\
&=&\int \frac{d^{2}\zeta }{\pi }\exp \{g\left[ \left( v_{1}+\zeta
_{1}\right) ^{2}+\left( v_{2}-\zeta _{2}\right) ^{2}\right] \}\left\vert
\zeta \right\rangle \left\langle \zeta \right\vert  \nonumber \\
&=&\int \frac{d^{2}\zeta }{\pi }\colon \exp \{-\left( 1-g\right) \left\vert
\zeta \right\vert ^{2}+\zeta \left( -K_{-}-i\Pi _{+}+gv\right) +\zeta ^{\ast
}\left( -K_{+}+i\Pi _{-}+gv^{\ast }\right)  \nonumber \\
&&+g|v|^{2}-\left( -K_{-}-i\Pi _{+}\right) \left( -K_{+}+i\Pi _{-}\right)
\}\colon  \nonumber \\
&=&\frac{1}{1-g}\colon \exp \{\frac{g}{1-g}\left( v^{\ast }-K_{+}+i\Pi
_{-}\right) \left( v-K_{-}-i\Pi _{+}\right) \}\colon .  \label{38}
\end{eqnarray}
$\ $

where $v=v_{1}+iv_{2}.$ Thus Eq. (\ref{37}) becomes (identifying $-1/\kappa $
in (\ref{37}) as $g$ in (\ref{38})$)$ $\ $
\begin{equation}
\int \frac{d^{2}\varepsilon }{\pi }\Delta _{h}\left( \gamma ,\varepsilon
,\kappa \right) =\frac{4}{\kappa }e^{-\frac{1}{\kappa }\left[ \left( \gamma
_{1}+\sqrt{\frac{2}{M\Omega }}p_{y}\right) ^{2}+\left( \gamma _{2}-\sqrt{%
\frac{2}{M\Omega }}p_{x}\right) ^{2}\right] },  \label{39}
\end{equation}%
so the another marginal distribution of (\ref{31}) is also a Gaussian
operator but with the factor $\frac{1}{\kappa }.$ It then follows from (\ref%
{39}) another marginal distribution of the Husimi function in "$\zeta -$%
direction"%
\begin{eqnarray}
\int \frac{d^{2}\varepsilon }{\pi }W_{h}\left( \gamma ,\varepsilon ;k\right)
&=&\left\langle \psi \right\vert \int \frac{d^{2}\varepsilon }{\pi }\Delta
_{h}\left( \gamma ,\varepsilon ;\kappa \right) \left\vert \psi \right\rangle
\nonumber \\
&=&\frac{4}{\kappa }\left\langle \psi \right\vert \int \frac{d^{2}\zeta }{%
\pi }e^{-\frac{1}{\kappa }\left[ \left( \gamma _{1}+\sqrt{\frac{2}{M\Omega }}%
p_{y}\right) ^{2}+\left( \gamma _{2}-\sqrt{\frac{2}{M\Omega }}p_{x}\right)
^{2}\right] }\left\vert \zeta \right\rangle \left\langle \zeta \right.
\left\vert \psi \right\rangle  \nonumber \\
&=&\frac{4}{\kappa }\left\langle \psi \right\vert \int \frac{d^{2}\zeta }{%
\pi }e^{-\frac{1}{\kappa }\left[ \left( \gamma _{1}+\zeta _{1}\right)
^{2}+\left( \gamma _{2}-\zeta _{2}\right) ^{2}\right] }\left\vert \zeta
\right\rangle \left\langle \zeta \right. \left\vert \psi \right\rangle
\nonumber \\
&=&\frac{4}{\kappa }\int \frac{d^{2}\zeta }{\pi }e^{-\frac{1}{\kappa }%
\left\vert \gamma ^{\ast }+\zeta \right\vert ^{2}}|\psi \left( \zeta \right)
|^{2},  \label{40}
\end{eqnarray}%
which is a Gaussian-broadened version of the quantal momentum probability
distribution $|\psi \left( \zeta \right) |^{2},$ (another Wigner marginal
distribution (comparing with Eq. (\ref{23})). Therefore, an
operator-representation theory which underlies the Husimi distribution of
electron in UMF is established, and the Husimi function's marginal
distributions are clear.

\section{The Husimi operator as a pure squeezed coherent state density
operator}

By noticing $\left\vert 00\right\rangle \left\langle 00\right\vert =:\exp
[-\Pi _{+}\Pi _{-}-K_{+}K_{-}]:$ we observe that the normally ordered form
of the Husimi operator $\Delta _{h}\left( \gamma ,\varepsilon ,\kappa
\right) $ in (\ref{31}) can be decomposed as

\begin{eqnarray}
&&\Delta _{h}\left( \gamma ,\varepsilon ;\kappa \right)  \nonumber \\
&=&\frac{4\kappa }{\left( 1+\kappa \right) ^{2}}\exp \{-\frac{1}{1+\kappa }%
[\kappa |\varepsilon |^{2}+|\gamma |^{2}-\left( \kappa \varepsilon +\gamma
\right) K_{+}+i\left( \kappa \varepsilon ^{\ast }-\gamma ^{\ast }\right) \Pi
_{+}-i\left( \kappa -1\right) \Pi _{+}K_{+}]\}  \nonumber \\
&&\times \colon \exp [-\Pi _{+}\Pi _{-}-K_{+}K_{-}]\colon \exp \{-\frac{1}{%
1+\kappa }[-\left( \kappa \varepsilon ^{\ast }+\gamma ^{\ast }\right)
K_{-}-i\left( \kappa \varepsilon -\gamma \right) \Pi _{-}+i\left( \kappa
-1\right) \Pi _{-}K_{-}\}  \nonumber \\
&=&\left\vert \gamma ,\varepsilon \right\rangle _{\kappa \kappa
}\left\langle \gamma ,\varepsilon \right\vert ,  \label{41}
\end{eqnarray}%
where we have defined the new state
\begin{eqnarray}
\left\vert \gamma ,\varepsilon \right\rangle _{\kappa } &=&\frac{2\sqrt{%
\kappa }}{1+\kappa }\exp \{-\frac{1}{1+\kappa }[\frac{\kappa |\varepsilon
|^{2}}{2}+\frac{|\gamma |^{2}}{2}  \label{42} \\
&&-\left( \kappa \varepsilon +\gamma \right) K_{+}+i\left( \kappa
\varepsilon ^{\ast }-\gamma ^{\ast }\right) \Pi _{+}-i\left( \kappa
-1\right) \Pi _{+}K_{+}]\}\left\vert 00\right\rangle .  \nonumber
\end{eqnarray}%
Thus the Husimi operator $\Delta _{h}\left( \lambda ,\zeta ,\kappa \right) $
is just the pure state density operator $\left\vert \gamma ,\varepsilon
\right\rangle _{\kappa \kappa }\left\langle \gamma ,\varepsilon \right\vert
, $ this is a remarkable result. It turns out that $\left\vert \gamma
,\varepsilon \right\rangle _{\kappa }$ is a two-mode squeezed canonical
coherent state because it obeys the eigenvector equations%
\begin{equation}
\left( K_{-}\cosh r+i\Pi _{+}\sinh r\right) \left\vert \gamma ,\varepsilon
\right\rangle _{\kappa }=\frac{\sqrt{\kappa }\varepsilon +\gamma /\kappa }{2}%
\left\vert \gamma ,\varepsilon \right\rangle _{\kappa }  \label{43}
\end{equation}%
and%
\begin{equation}
\left( \Pi _{-}\cosh r+iK_{+}\sinh r\right) \left\vert \gamma ,\varepsilon
\right\rangle _{\kappa }=i\frac{\gamma ^{\ast }/\sqrt{\kappa }-\sqrt{\kappa }%
\varepsilon ^{\ast }}{2}\left\vert \gamma ,\varepsilon \right\rangle
_{\kappa }  \label{44}
\end{equation}%
where $\frac{1-\kappa }{1+\kappa }\equiv \tanh r$ is a squeezing parameter, $%
e^{r}=\frac{1}{\sqrt{\kappa }},$ $\cosh r=\frac{1+\kappa }{2\sqrt{\kappa }}.$
The corresponding squeezing operator is%
\begin{equation}
S\left( r\right) \equiv e^{i(xp_{x}+yp_{y}-i)r}{}=\exp \left[ ir\left( \Pi
_{+}K_{+}+\Pi _{-}K_{-}\right) \right] ,  \label{45}
\end{equation}%
(For a review of general squeezed state theory in quantum optics we refer to
\cite{r24}). The disentangling of (\ref{45}) is
\begin{eqnarray}
S\left( r\right) &=&\sec hr\exp \left( i\Pi _{+}K_{+}\tanh r\right) \exp
[\left( K_{+}K_{-}+\Pi _{+}\Pi _{-}\right) \ln \sec hr]  \label{46} \\
&&\times \exp \left( i\Pi _{-}K_{-}\tanh r\right) .\text{ }  \nonumber
\end{eqnarray}%
From (\ref{46}), (\ref{14})-(\ref{15}) we derive%
\begin{eqnarray}
S^{-1}K_{-}S &=&K_{-}\cosh r+i\Pi _{+}\sinh r,\text{\ \ }S^{-1}\Pi _{-}S=\Pi
_{-}\cosh r+iK_{+}\sinh r,  \label{47} \\
S^{-1}K_{+}S &=&K_{+}\cosh r-i\Pi _{-}\sinh r,\text{\ \ }S^{-1}\Pi _{+}S=\Pi
_{+}\cosh r-iK_{-}\sinh r,  \nonumber
\end{eqnarray}%
and using (\ref{18}) and (\ref{27}) we have
\begin{equation}
S^{-1}xS=\sqrt{\kappa }x,\text{ }S^{-1}yS=\sqrt{\kappa }y,\text{ }
\label{48}
\end{equation}%
\begin{equation}
S^{-1}p_{x}S=p_{x}/\sqrt{\kappa },\text{ }S^{-1}p_{y}S=p_{y}/\sqrt{\kappa }.
\label{49}
\end{equation}%
In (\ref{19}) we see that $\lambda $ denotes the eigenvalue of electron's
coordinates, so $S\left( r\right) $ has a natural representation in $%
\left\langle \lambda \right\vert $ representation \cite{r25}%
\begin{equation}
S\left( r\right) =e^{-r}\int \frac{d^{2}\lambda }{\pi }\left\vert
e^{-r}\lambda \right\rangle \left\langle \lambda \right\vert ,\;e^{r}=\frac{1%
}{\sqrt{\kappa }},  \label{50}
\end{equation}%
from $\left\langle \lambda \right\vert \left. \lambda ^{\prime
}\right\rangle =\pi \delta ^{\left( 2\right) }\left( \lambda -\lambda
^{\prime }\right) $, $S\left( r\right) \left\vert \lambda \right\rangle =$ $%
e^{-r}\left\vert e^{-r}\lambda \right\rangle $ , so (\ref{50}) embodies
another merit of constructing the entangled state representation $\left\vert
\lambda \right\rangle $. From the eigenvalue equations (\ref{19}) we also
see that the eigenvalue of $x$ and $y$ varies with $B$, since $\sqrt{\frac{1%
}{M\Omega }}=\frac{1}{\sqrt{eB}}$, so the variation of the magnetic field
intensity $B$ is related to squeezing of electron's orbit track. Thus the
variation of Gaussian spatial width parameter $\sqrt{\kappa }$ can also be
interpreted as the change of magnetic field intensity $\sqrt{B}.$ From (\ref%
{43})-(\ref{44}) we notice that $\left\vert \gamma ,\varepsilon
\right\rangle _{\kappa }$ can be expressed as the result of the squeezing
operator operating on the state $\left\vert \gamma ,\varepsilon
\right\rangle ,$ i.e.%
\begin{equation}
\left\vert \gamma ,\varepsilon \right\rangle _{\kappa }=S^{-1}\left(
r\right) \left\vert \gamma ,\varepsilon \right\rangle ,  \label{51}
\end{equation}%
where%
\begin{eqnarray}
\left\vert \gamma ,\varepsilon \right\rangle &\equiv &\exp [-\frac{1}{4}%
\left( \kappa |\varepsilon |^{2}+|\gamma |^{2}/\kappa \right) +i\frac{\gamma
^{\ast }/\sqrt{\kappa }-\sqrt{\kappa }\varepsilon ^{\ast }}{2}\Pi _{+}
\label{52} \\
&&+\frac{\sqrt{\kappa }\varepsilon +\gamma /\sqrt{\kappa }}{2}%
K_{+}]\left\vert 00\right\rangle ,  \nonumber
\end{eqnarray}%
is a normalized two-mode coherent state \cite{r25} for an electron in UMF,
and we have dropped the inconsequential phase factor $\exp \{\frac{\kappa -1%
}{4(1+\kappa )}(\varepsilon ^{\ast }\gamma -\gamma ^{\ast }\varepsilon )\}$
in the result of calculating $S^{-1}\left( r\right) \left\vert \gamma
,\varepsilon \right\rangle .$

\section{Further explanation of the Husimi function}

Using (\ref{52}), (\ref{48}) and (\ref{18}) we see that in the state $%
\left\vert \gamma =0,\varepsilon =0\right\rangle _{\kappa }$ the variance of
electron's position $x$ is%
\begin{eqnarray}
\left( \Delta x\right) ^{2} &\equiv &\text{\ }_{\kappa }\left\langle
0,0\right\vert x^{2}\left\vert 0,0\right\rangle _{\kappa }-\left( _{\kappa
}\left\langle 0,0\right\vert x\left\vert 0,0\right\rangle _{\kappa }\right)
^{2}=\text{\ }\left\langle 0,0\right\vert S\left( r\right) x^{2}S^{-1}\left(
r\right) \left\vert 00\right\rangle  \nonumber \\
&=&\frac{1}{2M\Omega \kappa }\text{\ }\left\langle 00\right\vert \left(
K_{+}+K_{-}-i\Pi _{+}+i\Pi _{-}\right) ^{2}\left\vert 00\right\rangle =\frac{%
1}{M\Omega \kappa },  \label{53}
\end{eqnarray}%
while the variances of $p_{x}$ is
\begin{eqnarray}
\left( \Delta p_{x}\right) ^{2} &=&\left\langle 00\right\vert S\left(
r\right) p_{x}^{2}S^{-1}\left( r\right) \left\vert 00\right\rangle  \nonumber
\\
&=&\frac{\kappa M\Omega }{8}\left\langle 0,0\right\vert \left[ \Pi _{+}+\Pi
_{-}-iK_{+}+iK_{-}\right] ^{2}\left\vert 00\right\rangle =\frac{\kappa
M\Omega }{4}.  \label{54}
\end{eqnarray}%
On the other hand, $\left\vert \gamma ,\varepsilon \right\rangle _{\kappa }$
is complete
\begin{equation}
\frac{1}{4\pi ^{2}}\int d^{2}\varepsilon \int d^{2}\gamma \left\vert \gamma
,\varepsilon \right\rangle _{\kappa \kappa }\left\langle \gamma ,\varepsilon
\right\vert =1,  \label{55}
\end{equation}%
so the Husimi density
\begin{equation}
\left\langle \psi \right\vert \Delta _{h}\left( \gamma ,\varepsilon ,\kappa
\right) \left\vert \psi \right\rangle =\left\vert \left\langle \psi \right.
\left\vert \gamma ,\varepsilon \right\rangle _{\kappa }\right\vert ^{2}
\label{56}
\end{equation}%
is given by the projection of the wave function onto the squeezed coherent
states localized in phase space with a minimum product of the uncertainties%
\begin{equation}
\Delta p_{x}=\sqrt{\frac{\kappa M\Omega }{4}},\text{ \ }\Delta x=\sqrt{\frac{%
1}{M\Omega \kappa }},\text{\ }\Delta x\Delta p_{x}=\frac{1}{2}.  \label{57}
\end{equation}%
In this sense the Gaussian spatial width parameter $\kappa =\frac{2\Delta
p_{x}}{M\Omega \Delta x}=\frac{2\Delta p_{x}}{eB\Delta x}$ plays the role of
squeezing-parameter (note that in the units of $\hbar =c=1,$ $\sqrt{\frac{2}{%
eB}}$ is the magnetic length.) Further, using (\ref{41}) we can re-express
the marginal distribution (\ref{40}) of the Husimi function of electron's
quantum state $\left\vert \psi \right\rangle $ as
\begin{equation}
\int \frac{d^{2}\varepsilon }{\pi }W_{h}\left( \gamma ,\varepsilon ;k\right)
=\int \frac{d^{2}\varepsilon }{\pi }|_{\kappa }\left\langle \gamma
,\varepsilon \right. \left\vert \psi \right\rangle |^{2}.  \label{58}
\end{equation}%
We can also recast (\ref{36}) as%
\begin{equation}
\int \frac{d^{2}\gamma }{\pi }W_{h}\left( \gamma ,\varepsilon ;\kappa
\right) =\int \frac{d^{2}\gamma }{\pi }|_{\kappa }\left\langle \gamma
,\varepsilon \right. \left\vert \psi \right\rangle |^{2}.  \label{59}
\end{equation}%
Eqs. (\ref{58}) and (\ref{59}) indicate the relationship between probability
density of $\left\vert \psi \right\rangle $ in the $_{\kappa }\left\langle
\gamma ,\varepsilon \right\vert $ representation and those in the entangled
state $\left\langle \lambda \right\vert $ representation.

\section{Husimi functions of some electron's states}

Eq. (\ref{41}) brings great convenience to calculate Husimi functions of
various electron's states. Using the two-mode coherent state's completeness
relation \cite{r25}-\cite{r27}
\begin{equation}
\int \frac{d^{2}z_{1}d^{2}z_{2}}{\pi ^{2}}\left\vert
z_{1},z_{2}\right\rangle \left\langle z_{1},z_{2}\right\vert =1,  \label{60}
\end{equation}%
where
\begin{eqnarray}
\left\langle z_{1},z_{2}\right\vert &=&\left\langle 00\right\vert \exp \left[
-\frac{1}{2}\left( |z_{1}|^{2}+|z_{2}|^{2}\right) +z_{1}^{\ast }\Pi
_{-}+z_{2}^{\ast }K_{-}\right] ,  \label{61} \\
\left\langle z_{1},z_{2}\right\vert \Pi _{+} &=&\left\langle
z_{1},z_{2}\right\vert z_{1}^{\ast },\text{ \ \ }\left\langle
z_{1},z_{2}\right\vert K_{+}=\left\langle z_{1},z_{2}\right\vert z_{2}^{\ast
},  \nonumber
\end{eqnarray}%
and (\ref{42}) we immediately have

\begin{eqnarray}
\left\langle z_{1},z_{2}\right. \left\vert \gamma ,\varepsilon \right\rangle
_{\kappa } &=&\frac{2\sqrt{\kappa }}{1+\kappa }e^{-\left(
|z_{1}|^{2}+|z_{2}|^{2}\right) /2}  \nonumber \\
&&\times \exp \{-\frac{1}{1+\kappa }[\frac{\kappa |\varepsilon |^{2}}{2}+%
\frac{|\gamma |^{2}}{2}-\left( \kappa \varepsilon +\gamma \right)
z_{2}^{\ast }+i\left( \kappa \varepsilon ^{\ast }-\gamma ^{\ast }\right)
z_{1}^{\ast }-i\left( \kappa -1\right) z_{1}^{\ast }z_{2}^{\ast }]\}.
\label{62}
\end{eqnarray}
We further calculate the overlap

\begin{eqnarray}
_{\kappa }\left\langle \gamma ^{\prime },\varepsilon ^{\prime }\right.
\left\vert \gamma ,\varepsilon \right\rangle _{\kappa } &=&_{\kappa
}\left\langle \gamma ^{\prime },\varepsilon ^{\prime }\right\vert \int \frac{%
d^{2}z_{1}d^{2}z_{2}}{\pi ^{2}}\left\vert z_{1},z_{2}\right\rangle
\left\langle z_{1},z_{2}\right\vert \left. \gamma ,\varepsilon \right\rangle
_{\kappa }  \nonumber \\
&=&\frac{4\kappa }{\left( 1+\kappa \right) ^{2}}\exp \{-\frac{1}{2\left(
1+\kappa \right) }[\kappa |\varepsilon |^{2}+|\gamma |^{2}+\kappa
|\varepsilon ^{\prime }|^{2}+|\gamma ^{\prime }|^{2}]\}  \nonumber \\
&&\times \int \frac{d^{2}z_{1}d^{2}z_{2}}{\pi ^{2}}\exp
\{-|z_{1}|^{2}-|z_{2}|^{2}-\frac{1}{1+\kappa }[-\left( \kappa \varepsilon
^{\prime \ast }+\gamma ^{\prime \ast }\right) z_{2}-i\left( \kappa
\varepsilon ^{\prime }-\gamma ^{\prime }\right) z_{1}  \nonumber \\
&&+i\left( \kappa -1\right) z_{1}z_{2}-\left( \kappa \varepsilon +\gamma
\right) z_{2}^{\ast }+i\left( \kappa \varepsilon ^{\ast }-\gamma ^{\ast
}\right) z_{1}^{\ast }-i\left( \kappa -1\right) z_{1}^{\ast }z_{2}^{\ast }]\}
\nonumber \\
&=&\exp \{-\frac{\kappa }{4}\left\vert \varepsilon ^{\prime }-\varepsilon
\right\vert ^{2}-\frac{\left\vert \gamma ^{\prime }-\gamma \right\vert ^{2}}{%
4\kappa }+\frac{1}{4}\left( \gamma ^{\prime \ast }\varepsilon -\varepsilon
^{\ast }\gamma ^{\prime }+\gamma \varepsilon ^{\prime \ast }-\varepsilon
^{\prime }\gamma ^{\ast }\right)  \nonumber \\
&&+\frac{\kappa -1}{4\left( 1+\kappa \right) }\left( \varepsilon ^{\prime
\ast }\gamma ^{\prime }-\varepsilon ^{\prime }\gamma ^{\prime \ast
}+\varepsilon \gamma ^{\ast }-\varepsilon ^{\ast }\gamma \right) \},
\label{63}
\end{eqnarray}

where the third and fourth terms in the last exponential are all pure
imaginary, so we immediately obtain the Husimi function of $\left\vert
\varepsilon ^{\prime },\gamma ^{\prime }\right\rangle _{k},$
\begin{eqnarray}
_{\kappa }\left\langle \varepsilon ^{\prime },\gamma ^{\prime }\right\vert
\Delta _{h}\left( \varepsilon ,\gamma ,\kappa \right) \left\vert \varepsilon
^{\prime },\gamma ^{\prime }\right\rangle _{\kappa } &=&\left\vert _{\kappa
}\left\langle \varepsilon ,\gamma \right\vert \left. \varepsilon ^{\prime
},\gamma ^{\prime }\right\rangle _{\kappa }\right\vert ^{2}  \label{64} \\
&=&\exp \left[ -\frac{\kappa }{2}\left\vert \varepsilon ^{\prime
}-\varepsilon \right\vert ^{2}-\frac{\left\vert \gamma ^{\prime }-\gamma
\right\vert ^{2}}{2\kappa }\right] ,  \nonumber
\end{eqnarray}%
which is also a Gaussian broadened function. Further, using (\ref{50})-(\ref%
{52}) and (\ref{12}) we have

\begin{eqnarray}
\left\langle \lambda \right\vert \left. \gamma ,\varepsilon \right\rangle
_{\kappa } &=&\left\langle \lambda \right\vert S^{-1}\left( r\right)
\left\vert \gamma ,\varepsilon \right\rangle =\sqrt{\kappa }\left\langle
\sqrt{\kappa }\lambda \right\vert \left. \gamma ,\varepsilon \right\rangle
\nonumber \\
&=&\sqrt{\kappa }\exp \{-\frac{1}{4}\left( \kappa |\varepsilon |^{2}+|\gamma
|^{2}/\kappa \right) -\frac{1}{2}\kappa \left\vert \lambda \right\vert
^{2}-\lambda ^{\ast }\frac{\gamma ^{\ast }-\kappa \varepsilon ^{\ast }}{2}
\nonumber \\
&&+\lambda \frac{\kappa \varepsilon +\gamma }{2}+\frac{\gamma ^{\ast }/\sqrt{%
\kappa }-\sqrt{\kappa }\varepsilon ^{\ast }}{2}\frac{\gamma /\sqrt{\kappa }+%
\sqrt{\kappa }\varepsilon }{2}\}  \nonumber \\
&=&\sqrt{\kappa }\exp \{\left[ -\frac{1}{2}\kappa \left( |\varepsilon
|^{2}+\left\vert \lambda \right\vert ^{2}\right) +\kappa \func{Re}\left(
\lambda \varepsilon \right) +i\func{Im}\lambda \gamma +i\func{Im}\varepsilon
\gamma ^{\ast }\right] ,  \label{65}
\end{eqnarray}%
so the Husimi function of the electron's coordinate eigenstate $\left\vert
\lambda \right\rangle $ is%
\begin{equation}
\left\langle \lambda \right\vert \Delta _{h}\left( \gamma ,\varepsilon
;\kappa \right) \left\vert \lambda \right\rangle =\kappa |\left\langle \sqrt{%
\kappa }\lambda \right\vert \left. \gamma ,\varepsilon \right\rangle
|^{2}=\kappa \exp \{-\kappa |\lambda -\varepsilon ^{\ast }|^{2}\},
\label{66}
\end{equation}%
which is a Gaussian. This is in sharply contrast with the Wigner function of
$\left\vert \lambda \right\rangle $ which can be calculated by using (\ref%
{11})%
\begin{eqnarray}
\left\langle \lambda \right\vert \Delta _{B}\left( \gamma ,\varepsilon
\right) \left\vert \lambda \right\rangle &\equiv &\left\langle \lambda
\right\vert \int \frac{d^{2}\lambda ^{\prime }}{\pi ^{3}}\left\vert
\varepsilon ^{\ast }-\lambda ^{\prime }\right\rangle \left\langle
\varepsilon ^{\ast }+\lambda ^{\prime }\right\vert e^{\gamma ^{\ast }\lambda
^{\prime \ast }-\gamma \lambda ^{\prime }}\left\vert \lambda \right\rangle
\label{67} \\
&=&\int \frac{d^{2}\lambda ^{\prime }}{\pi }\delta ^{\left( 2\right) }\left(
\lambda -\varepsilon ^{\ast }+\lambda ^{\prime }\right) \delta ^{\left(
2\right) }\left( \lambda -\varepsilon ^{\ast }-\lambda ^{\prime }\right)
e^{\gamma ^{\ast }\lambda ^{\prime \ast }-\gamma \lambda ^{\prime }}
\nonumber \\
&=&\frac{1}{4\pi }\delta ^{\left( 2\right) }\left( \lambda -\varepsilon
^{\ast }\right) .  \nonumber
\end{eqnarray}%
From (\ref{11}) we see $\varepsilon =\chi -i\sigma ^{\ast }=\sqrt{\frac{%
M\Omega }{2}}\left( q_{1}+iq_{2}\right) ,$ so
\begin{equation}
\left\langle \lambda \right\vert \Delta _{B}\left( \gamma ,\varepsilon
\right) \left\vert \lambda \right\rangle =\frac{1}{4\pi }\delta \left(
\lambda _{1}-\sqrt{\frac{M\Omega }{2}}q_{1}\right) \delta \left( \lambda
_{2}+\sqrt{\frac{M\Omega }{2}}q_{2}\right) ,  \label{68}
\end{equation}%
which is in consistent with Eq. (\ref{18}). Comparing (\ref{66}) and (\ref%
{67}) and recall the limiting Gaussian-form of Delta function we can see
again that Husimi function is the Gaussian-broadened version of Wigner
function$.$ Next we consider a Landau state,%
\begin{equation}
\left\vert n,m\right\rangle =\frac{\Pi _{+}^{n}K_{+}^{m}}{\sqrt{n!m!}}%
\left\vert 00\right\rangle =\frac{1}{\sqrt{n!m!}}\frac{\partial ^{n}}{%
\partial z_{1}^{n}}\frac{\partial ^{m}}{\partial z_{2}^{m}}e^{z_{1}\Pi
_{+}}e^{z_{2}K_{+}}\left\vert 00\right\rangle |_{z_{1}=z_{2}=0}  \label{69}
\end{equation}%
where $n,m=0,1,2,...,$ from (\ref{62}) we know

\begin{eqnarray}
\left\langle n,m\right\vert \left. \gamma ,\varepsilon \right\rangle
_{\kappa } &=&\frac{1}{\sqrt{n!m!}}\frac{\partial ^{n}}{\partial
z_{1}^{^{\ast }n}}\frac{\partial ^{m}}{\partial z_{2}^{\ast m}}\text{ }%
\left\langle z_{1},z_{2}\right\vert \left. \gamma ,\varepsilon \right\rangle
_{\kappa }|_{z_{1}^{\ast }=z_{2}^{\ast }=0}  \label{70} \\
&=&\frac{1}{\sqrt{n!m!}}\left\langle z_{1},z_{2}\right\vert \frac{2\sqrt{%
\kappa }}{1+\kappa }\exp \{-\frac{1}{1+\kappa }[\frac{\kappa |\varepsilon
|^{2}}{2}+\frac{|\gamma |^{2}}{2}]\}\frac{\partial ^{n}}{\partial
z_{1}^{^{\ast }n}}\frac{\partial ^{m}}{\partial z_{2}^{\ast m}}  \nonumber \\
&&\times \text{ }\exp \{\frac{-1}{1+\kappa }[-\left( \kappa \varepsilon
+\gamma \right) z_{2}^{\ast }+i\left( \kappa \varepsilon ^{\ast }-\gamma
^{\ast }\right) z_{1}^{\ast }-i\left( \kappa -1\right) z_{1}^{\ast
}z_{2}^{\ast }]\}\left\vert 00\right\rangle |_{z_{1}^{\ast }=z_{2}^{\ast }=0}
\nonumber \\
&=&\frac{i^{m}}{\sqrt{n!m!}}\frac{2\sqrt{\kappa }}{1+\kappa }\exp \{-\frac{1%
}{1+\kappa }[\frac{\kappa |\varepsilon |^{2}}{2}+\frac{|\gamma |^{2}}{2}]\}
\nonumber \\
&&\times \left( \frac{1-\kappa }{1+\kappa }\right) ^{\left( m+n\right)
/2}H_{m,n}\left( \frac{-\left( \kappa \varepsilon +\gamma \right) }{\sqrt{%
\kappa ^{2}-1}},\frac{-\left( \kappa \varepsilon ^{\ast }-\gamma ^{\ast
}\right) }{\sqrt{\kappa ^{2}-1}}\right) .  \nonumber
\end{eqnarray}%
where $H_{m,n}$ is two-variable Hermite polynomial\cite{r28} whose
definition is%
\begin{equation}
H_{m,n}(x,y)=\sum_{l=0}^{\min (m,n)}\frac{m!n!(-1)^{l}}{l!(m-l)!(n-l)!}%
x^{m-l}y^{n-l},  \label{71}
\end{equation}%
(which is not a direct product of two independent single-variable Hermite
polynomials). The generating function of $H_{m,n}(x,y)$ is
\begin{equation}
\sum_{m,n=0}^{\infty }\frac{z^{m}z^{\prime n}}{m!n!}H_{m,n}(x,y)=\exp
\{-zz^{\prime }+zx+z^{\prime }y\},  \label{72}
\end{equation}%
so%
\begin{equation}
H_{m,n}\left( x,y\right) =\frac{\partial ^{m}}{\partial z^{m}}\frac{\partial
^{n}}{\partial z^{\prime n}}e^{-zz^{\prime }+zx+z^{\prime }y}|_{z=z^{\prime
}=0}  \label{73}
\end{equation}%
Thus the Husimi function of $\left\vert n,m\right\rangle $ is%
\begin{eqnarray}
\left\langle n,m\right\vert \Delta _{h}\left( \gamma ,\varepsilon ;\kappa
\right) \left\vert n,m\right\rangle &=&|\left\langle n,m\right\vert \left.
\gamma ,\varepsilon \right\rangle _{\kappa }|^{2}  \nonumber \\
&=&\frac{1}{n!m!}\frac{4\kappa }{\left( 1+\kappa \right) ^{2}}\left( \frac{%
1-\kappa }{1+\kappa }\right) ^{n+m}\exp \left( -\frac{\kappa |\varepsilon
|^{2}+|\gamma |^{2}}{1+\kappa }\right)  \nonumber \\
&&\times \left\vert H_{m,n}\left( \frac{-\left( \kappa \varepsilon +\gamma
\right) }{\sqrt{\kappa ^{2}-1}},\frac{-\left( \kappa \varepsilon ^{\ast
}-\gamma ^{\ast }\right) }{\sqrt{\kappa ^{2}-1}}\right) \right\vert ^{2}.
\label{74}
\end{eqnarray}

\section{\protect\bigskip Squeezing of Husimi function by variation of
magnetic field}

In (\ref{48}) we have mentioned that the variation of magnetic field
intensity may cause squeezing of orbit track of electron's motion. Let the
corresponding squeezing operator is $S\left( \mu \right) ,$ in the $%
\left\vert \lambda \right\rangle $ representation it is expressed by (see
Appendix)%
\begin{equation}
S\left( \mu \right) =\int \frac{d^{2}\lambda }{\pi \mu }\left\vert \lambda
/\mu \right\rangle \left\langle \lambda \right\vert .  \label{75}
\end{equation}%
Under the squeezing transform the Wigner operator changes
\begin{equation}
S\left( \mu \right) \Delta _{B}\left( \gamma ,\varepsilon \right)
S^{-1}\left( \mu \right) =\int \frac{d^{2}\lambda }{\pi ^{3}}\left\vert
\frac{\varepsilon ^{\ast }}{\mu }-\lambda \right\rangle \left\langle \frac{%
\varepsilon ^{\ast }}{\mu }+\lambda \right\vert e^{\mu \left( \gamma ^{\ast
}\lambda ^{\ast }-\gamma \lambda \right) }=\Delta _{B}\left( \mu \gamma
,\varepsilon /\mu \right) .  \label{76}
\end{equation}%
From (\ref{30}) we see that the Husimi operator becomes
\begin{eqnarray}
S\left( \mu \right) \Delta _{h}\left( \gamma ,\varepsilon ;k\right)
S^{-1}\left( \mu \right) &=&4\int d^{2}\gamma ^{\prime }d^{2}\varepsilon
^{\prime }\Delta _{B}\left( \mu \gamma ^{\prime },\varepsilon ^{\prime }/\mu
\right) \exp \left[ -\kappa \left\vert \varepsilon -\varepsilon ^{\prime
}\right\vert ^{2}-\frac{\left\vert \gamma -\gamma ^{\prime }\right\vert ^{2}%
}{\kappa }\right]  \nonumber \\
&=&\Delta _{h}\left( \mu \gamma ,\varepsilon /\mu ;k\mu ^{2}\right) .
\label{77}
\end{eqnarray}%
we again see the squeezing parameter $\mu $ is equivalent to the Gaussian
broaden parameter $1/\sqrt{k}.$ (\ref{77}) and (\ref{41}) indicates
\begin{equation}
S\left( \mu \right) \left\vert \gamma ,\varepsilon \right\rangle _{\kappa
}=\left\vert \mu \gamma ,\varepsilon /\mu \right\rangle _{\kappa \mu ^{2}}.
\label{78}
\end{equation}%
From (\ref{31}) we see the Husimi function of the lowest Landau state is
\begin{equation}
\left\langle 00\right\vert \Delta _{h}\left( \gamma ,\varepsilon ;k\right)
\left\vert 00\right\rangle =\frac{4\kappa }{\left( 1+\kappa \right) ^{2}}%
\exp \{-\frac{\kappa }{1+\kappa }|\varepsilon |^{2}-\frac{1}{1+\kappa }%
|\gamma |^{2}\}.  \label{79}
\end{equation}%
Using (\ref{41}), (\ref{51}), (\ref{77}) and (\ref{79}) we immediately
obtain the Husimi function of squeezed Landau vacuum state,
\begin{eqnarray}
&&\left\langle 00\right\vert S\left( \mu \right) \Delta _{h}\left( \gamma
,\varepsilon ;\kappa \right) S^{-1}\left( \mu \right) \left\vert
00\right\rangle =\left\langle 00\right\vert \Delta _{h}\left( \mu \gamma
,\varepsilon /\mu ;\kappa \mu ^{2}\right) \left\vert 00\right\rangle
\label{80} \\
&=&\frac{4\kappa \mu ^{2}}{\left( 1+\kappa \mu ^{2}\right) ^{2}}\exp \{-%
\frac{\kappa }{\kappa \mu ^{2}+1}\left\vert \varepsilon \right\vert ^{2}-%
\frac{\mu ^{2}}{\kappa \mu ^{2}+1}\left\vert \gamma \right\vert ^{2}\}.
\nonumber
\end{eqnarray}

In summary, for the first time we have introduced the Husimi operator $%
\Delta _{h}\left( \gamma ,\varepsilon ;\kappa \right) $ for electron in UMF,
and shown $\Delta _{h}\left( \lambda ,\zeta ,\kappa \right) =\left\vert
\lambda ,\zeta \right\rangle _{\kappa \kappa }\left\langle \lambda ,\zeta
\right\vert ,$ i.e. the Husimi operator actually is a pure squeezed coherent
state projector. The normally ordered form of Husimi operator are also
derived which provides us with an operator version to examine various
properties of the Husimi distribution. We have in many ways demonstrated
that \ Husimi (marginal) distributions are Gaussian-broadened version of the
Wigner (marginal) distributions. Throughout the paper we have fully employed
the technique of integration within an ordered product of operators and the
entangled state representation, each of them seems an efficient method for
studying quantum statistical physics \cite{r30}.

\section{Appendix}

Using (\ref{12}) the IWOP technique we can derive $S\left( \mu \right)
^{\prime }$s normal ordering \cite{r29},

\begin{eqnarray*}
S\left( \mu \right) &=&\int \frac{d^{2}\lambda }{\pi \mu }\left\vert \lambda
/\mu \right\rangle \left\langle \lambda \right\vert =\int \frac{d^{2}\lambda
}{\pi \mu }\colon \exp \{-\frac{1}{2}\left\vert \lambda \right\vert
^{2}\left( 1+\frac{1}{\mu ^{2}}\right) +\lambda \left( K_{-}-i\frac{\Pi _{+}%
}{\mu }\right) \\
&&+\lambda ^{\ast }\left( \frac{K_{+}}{\mu }+i\Pi _{-}\right) +i\Pi
_{+}K_{+}-i\Pi _{-}K_{-}-K_{+}K_{-}-\Pi _{+}\Pi _{-}\}\colon \\
&=&\frac{2\mu }{1+\mu ^{2}}\colon \exp \left[ \frac{2\mu ^{2}}{1+\mu ^{2}}%
\left( K_{-}-i\frac{\Pi _{+}}{\mu }\right) \left( \frac{K_{+}}{\mu }+i\Pi
_{-}\right) -\left( K_{-}-i\Pi _{+}\right) \left( K_{+}+i\Pi _{-}\right) %
\right] \colon \\
&=&\sec hf\exp \left( i\Pi _{+}K_{+}\tanh f\right) \exp [\left(
K_{+}K_{-}+\Pi _{+}\Pi _{-}\right) \ln \sec hf]\exp \left( i\Pi
_{-}K_{-}\tanh f\right) ,
\end{eqnarray*}%
where $\mu =e^{f},$ we can say that the classical dilation $\lambda
\rightarrow \tfrac{\lambda }{\mu }$ maps into the squeezing operator $%
S\left( \mu \right) .$(\ref{75}) again realizes Dirac's statement that the
symbolic method can \textquotedblleft express the physical law in a neat and
concise way\textquotedblright .

\end{document}